\documentclass[12pt]{article}
\usepackage{a4wide}
\usepackage{epsfig}
\usepackage{amsmath}
\usepackage{latexsym}
\usepackage{multirow}

\newcommand{\half}{\textstyle\frac{1}{2}}
\newcommand{\fourth}{\textstyle\frac{1}{4}}
\newcommand{\threehalf}{\textstyle\frac{3}{2}}

\newcommand{\GF}{G_{\rm F}}

\def\gsim{\mathrel{\rlap{\raise 2.5pt \hbox{$>$}}\lower 2.5pt
\hbox{$\sim$}}}
\def\lsim{\mathrel{\rlap{\raise 2.5pt \hbox{$<$}}\lower 2.5pt
\hbox{$\sim$}}}

\def\Month{\ifcase\month\or
January\or February\or March\or April\or May\or June\or
July\or August\or September\or October\or November\or December\fi}
\renewcommand{\theequation}{\thesection.\arabic{equation}}

\catcode`@=11
\def\citer{\@ifnextchar [{\@tempswatrue\@citexr}{\@tempswafalse\@citexr[]}}

\def\@citexr[#1]#2{\if@filesw\immediate
  \write\@auxout{\string\citation{#2}}\fi
  \def\@citea{}\@cite{\@for\@citeb:=#2\do
    {\@citea\def\@citea{--\penalty\@m}\@ifundefined
       {b@\@citeb}{{\bf ?}\@warning
       {Citation `\@citeb' on page \thepage \space undefined}}%
\hbox{\csname b@\@citeb\endcsname}}}{#1}}
\catcode`@=12

\begin{document}
 \thispagestyle{empty}
 \begin{flushright}
 {CERN-TH/2000-097} \\[2mm]
\end{flushright}
 \vspace*{2.0cm}
 \begin{center}
 {\bf \Large
 Mikheyev-Smirnov-Wolfenstein Effect\\[4mm]
 for Linear Electron Density}
 \end{center}
 \vspace{1cm}
 \begin{center}
Harry Lehmann\footnote{Deceased 22 November 1998.} \\
II.\ Institut f\"ur Theoretische Physik der Universit\"at
Hamburg, Hamburg, Germany \\
\vspace{8mm}
Per Osland\footnote{This work was supported in part by the Research
Council of Norway.} \\
Department of Physics, University of Bergen, \\
      All\'{e}gaten 55, N-5007 Bergen, Norway 
\vspace{8mm}
and \\
Tai Tsun Wu\footnote{Work supported in part by the United
States Department of Energy under Grant No.\ DE-FG02-84ER40158.} \\
Gordon McKay Laboratory, Harvard University, Cambridge,
Massachusetts 02138, and\\
Theoretical Physics Division, CERN, CH-1211 Geneva 23, Switzerland
 \vspace{1cm}
\end{center}

\begin{abstract}
When the electron density is a linear function of distance,
it is known that the MSW equations for two neutrino species can be
solved in terms of known functions.
It is shown here that more generally, for any number of neutrino species,
these MSW equations can be solved exactly in terms of single integrals.
While these integrals cannot be expressed in terms of known functions,
some of their simple properties are obtained.
Application to the solar neutrino problem is briefly discussed.
\end{abstract}

\vfill
\begin{flushleft}
 {CERN-TH/2000-097} \\[2mm]
 {March 2000}
\end{flushleft}
\newpage

\setcounter{footnote}{0}


\section{Introduction}
\setcounter{equation}{0}
In studying the Mikheyev-Smirnov-Wolfenstein (MSW) effect \cite{MSW}
due to the coherent forward scattering of neutrinos by electrons
in matter, it is often instructive to consider first special cases
where the electron density is taken to be a simple function of 
distance. It is the purpose of the present paper to investigate
perhaps the simplest case: the case where the electron density
is a linear function of distance.

The problem of the linear electron density is formulated in Sec.~2.
The case of two neutrino species has a long history \cite{two-nu},
and the solution, as reviewed in Sec.~3, can be expressed in terms of
parabolic cylinder functions---see, for example, Chapter~VIII of 
\cite{Bateman-2}, or equivalently confluent hypergeometric functions.
However, the solution in this form is specific to the case of
two neutrino species, and is not convenient for generalizations
to more neutrino species.
Physically, this generalization is essential because there are
at least three types of neutrinos.
Therefore, Sec.~4 is devoted to treating in a different way
the MSW differential equations for linear electron density and
two neutrino species.
On the one hand, this alternative method must lead to the same solutions
as those in Sec.~3; on the other hand, this new treatment can be readily
generalized to any number of neutrino species.
This case of linear electron density but any number of neutrino species
forms the main content of the present paper,
and various aspects of this case are treated in Sec.~5 and Sec.~6.
\section{Formulation of the Problem}
\setcounter{equation}{0}
Let there be $N$ types of neutrinos, denoted by $\nu_1$, $\nu_2$,
$\ldots$ $\nu_N$, where $\nu_1$ is the neutrino of the first generation,
i.e., the one that forms the SU(2) doublet with the electron.
It is assumed that $\nu_1$ is the only neutrino which interacts 
differently with the electron because of the exchange of the intermediate
boson $W$, while the others neutrinos $\nu_2$, $\nu_3$,
$\ldots$ $\nu_N$ all have the same interaction with the electron.
Thus, the neutrino mass matrix $M$ \cite{Rosen} is an $N\times N$
matrix. The eigenvalues of $M$ give the $N$ neutrino masses $\mu$.

In analyzing the MSW effect, the neutrino masses are usually taken
to be much smaller than the momentum $p$ of the neutrino.
Under this assumption, because
\begin{equation}
(p^2+\mu^2)^{1/2}\sim p+\frac{1}{2p}\,\mu^2,
\end{equation}
it is $M^2$ that enters in the differential equation for the MSW
effect. Let $\Psi(x)$ be the $N$-component neutrino wave function,
then this differential equation is
\begin{equation}
\label{Eq:MSW-d.e.}
i\frac{d}{dx}\,\Psi(x)=\left[W(x)+\frac{1}{2p}\,M^2\right]\Psi(x)
\end{equation}
where $W(x)$ is an $N\times N$ matrix whose only non-zero element is
\begin{equation}
\label{Eq:W11}
\left[W(x)\right]_{11}=\sqrt{2}\,\GF N_e(x),
\end{equation}
with $\GF$ the Fermi weak-interaction constant and $N_e(x)$
the electron density at the point $x$.

The terminology ``linear electron density'' is used to mean that 
$N_e(x)$ is a linear function of $x$.

Since $N_e(x)$ is the density of electrons, it cannot be negative.
Therefore, the MSW differential equation (\ref{Eq:MSW-d.e.})
is physically meaningful only for the half-line of $x$ where
$N_e(x)\ge0$. On the other hand, when the neutrino or the electron,
but not both, is replaced by its antiparticle, the quantity
$\left[W(x)\right]_{11}$ of Eq.~(\ref{Eq:W11}) changes sign.
Therefore, the complementary half-line of $x$ describes this slightly
different physical situation.
For this reason, Eq.~(\ref{Eq:MSW-d.e.}) is to be studied for the
entire range of $x$, from $-\infty$ to $+\infty$.

For the present case of the linear electron density, 
Eq.~(\ref{Eq:MSW-d.e.}) can be reduced, for a given value of $p$, 
to the dimensionless standard form
\begin{equation}
\label{Eq:psi-Schr}
i\frac{d}{dt}\,\psi(t)=A(t)\psi(x),
\end{equation}
where
\begin{equation}
\label{Eq:psi}
\psi(t)=
\begin{bmatrix}
\psi_1(t) \\[2pt] \psi_2(t) \\[2pt] \psi_3(t) \\[2pt] \vdots \\[2pt] 
\psi_N(t)
\end{bmatrix}
\end{equation}
and
\begin{equation}
\label{Eq:A}
A(t)=
\begin{bmatrix}
-t\,  & a_2\, & a_3\, & \hdots & a_N \\[1pt]
a_2   & b_2 & 0   & \hdots & 0 \\[1pt]
a_3   & 0   & b_3 & \hdots & 0 \\[1pt]
\vdots & \vdots   & \vdots &  & \vdots \\[1pt]
a_N & 0   & 0   & \hdots & b_N
\end{bmatrix}.
\end{equation}
This is accomplished as follows.

(i)
To change the independent variable from $x$ to $t$, there is a shift
in origin and a rescaling with possibly a reversal of sign.

(ii)
To change the dependent variable from $\Psi$ to $\psi$, there is
a rotation in the second to $N$th component and an introduction
of exponential factors with possibly some minus signs.

Furthermore, from (i) and (ii), the elements of the matrix $A(t)$
can be chosen to satisfy the conditions
\begin{equation}
\label{Eq:sum-b-zero}
\sum_{j=2}^N b_j=0,
\end{equation}
\begin{equation}
b_2\le b_3\le b_4\le\hdots\le b_{N-1}\le b_N,
\end{equation}
and
\begin{equation}
a_j\ge0 \qquad \text{for} \quad j=2,3,4,\hdots N.
\end{equation}

Consider the following special cases.

(a)
If, for some $j$, say $j_0$, $a_{j_0}=0$, then it is seen from
Eqs.~(\ref{Eq:psi}) and (\ref{Eq:A}) that $\psi_{j_0}$ is decoupled
from the other $\psi_{j}$'s.
Thus, this special case of $N$ types of neutrinos is reduced
to a problem with $N-1$ types of neutrinos.

(b)
If, again for some $j$, say $j_0$, $b_{j_0}=b_{j_0+1}$, then a rotation
can be carried out between $\psi_{j_0}$ and $\psi_{j_0+1}$ such that,
after this additional rotation, the new $a_{j_0}$ is zero.
Thus, this special case of $b_{j_0}=b_{j_0+1}$ can be reduced to the
above case of $a_{j_0}=0$, and hence again this second special case
of $N$ types of neutrinos is reduced to a problem with $N-1$ types 
of neutrinos.

It is therefore sufficient to study the ordinary differential equation
(\ref{Eq:psi-Schr}) with Eqs.~(\ref{Eq:psi}) and (\ref{Eq:A}) under 
the condition (\ref{Eq:sum-b-zero}) together with
\begin{equation}
\label{Eq:b-ordered}
b_2< b_3 < b_4 < \hdots < b_{N-1} < b_N
\end{equation}
and
\begin{equation}
\label{Eq:a-positive}
a_j>0 \qquad \text{for}\quad j=2,3,4,\hdots N.
\end{equation}
In view of the inequality (\ref{Eq:b-ordered}), it turns out to be
convenient to define symbolically
\begin{equation}
\label{Eq:b-extras}
b_1=-\infty \qquad \text{and} \quad b_{N+1}=+\infty.
\end{equation}
\section{Case $N=2$}
\setcounter{equation}{0}
Let us review first the well-known case of the MSW effect for two types
of neutrinos \cite{MSW,two-nu}. 
By Eqs.~(\ref{Eq:MSW-d.e.})--(\ref{Eq:sum-b-zero}), 
the MSW equations are
\begin{equation}
i\frac{d}{dt}
\begin{bmatrix}
\psi_1(t) \\ \psi_2(t)
\end{bmatrix}
=\begin{bmatrix}
 -t & a_2 \\
a_2 & 0
\end{bmatrix}
\begin{bmatrix}
\psi_1(t) \\ \psi_2(t)
\end{bmatrix},
\end{equation}
or more explicitly
\begin{eqnarray}
\label{Eq:MSW-2-nu}
i\frac{d}{dt}\psi_1(t)
&=&-t\psi_1(t)+a_2\psi_2(t), \\
\label{Eq:MSW-2-nu-last}
i\frac{d}{dt}\psi_2(t)
&=&a_2\psi_1(t),
\end{eqnarray}
with $a_2>0$. 
A second-order ordinary differential equation for $\psi_1(t)$ is obtained
by applying $d/dt$ to Eq.~(\ref{Eq:MSW-2-nu}) 
and using Eq.~(\ref{Eq:MSW-2-nu-last}):
\begin{equation}
\label{Eq:MSW-2-nu-other}
\frac{d^2\psi_1(t)}{dt^2}-it\,\frac{d\psi_1(t)}{dt}
+(a_2^2-i)\psi_1(t)=0.
\end{equation}

In order to remove the first-derivative term, let
\begin{equation}
\psi_1(t)=e^{it^2/4}\,\phi_1(t)
\end{equation}
Then the equation for $\phi_1(t)$ is
\begin{equation}
\label{Eq:MSW-2-nu-second-o}
\frac{d^2\phi_1(t)}{dt^2}+(\fourth t^2+a_2^2-\half i)\phi_1(t)=0
\end{equation}
Two linearly independent solutions of this 
Eq.~(\ref{Eq:MSW-2-nu-second-o}) are the parabolic cylinder functions
\cite{Bateman-2}
\begin{equation}
D_\rho(\pm e^{i\pi/4}\,t),
\end{equation}
where
\begin{equation}
\rho=-ia_2^2-1.
\end{equation}

Parabolic cylinder functions are special cases of the confluent
hypergeometric function \cite{Bateman-1}, the relation being
\begin{equation}
D_\rho(z)=2^{(\rho-1)/2}\, e^{-z^2/4}\, z\,
\Psi(\half-\half\rho,\threehalf;\half z^2).
\end{equation}
Since the confluent hypergeometric functions $\Psi$ and $\Phi$
satisfy the same second-order differential equation, the general
solution of Eq.~(\ref{Eq:MSW-2-nu-second-o}) is
\begin{equation}
\psi_1(t)
=t\bigl[C\,\Phi(1+\half ia_2^2,\threehalf;\half i t^2)
+C'\,\Psi(1+\half ia_2^2,\threehalf;\half i t^2)\bigr].
\end{equation}
This is one convenient form for the solution for $N=2$.
\section{Case $N=2$---an Alternative Approach}
\setcounter{equation}{0}
In the existing treatment in the literature for linear electron density
and two types of neutrinos \cite{MSW,two-nu} as reviewed in Sec.~3,
the crucial step is to recognize that the second-order differential 
equation (\ref{Eq:MSW-2-nu-other}) can be solved exactly in terms of
known higher transcendental functions, either parabolic cylinder
functions or confluent hypergeometric functions.
More generally, for $N$ types of neutrinos, the corresponding
differential equation is of $N$th order.
Even for $N=3$, the third-order differential equation is not one
for any well-known transcendental function.
Therefore, in order to be able to generalize the treatment of $N=2$
to larger values of $N$, we must recast the solution of Sec.~3 so that
parabolic cylinder functions and confluent hypergeometric functions
do not play an essential role.

A useful question to ask is the following:
In what way is the linear electron density especially simple?
The answer must be sought in Eq.~(\ref{Eq:A}), from which it is seen
that the independent variable $t$ appears only in one matrix element,
and furthermore, it appears only linearly in that element.
This implies that, if Fourier transform is applied to the differential
equation (\ref{Eq:psi-Schr}), the differentiation with respect
to the Fourier-transform variable appears only once.
Hence it is expected that an explicit expression can be obtained
for the Fourier transform of $\psi$.

Let
\begin{equation}
\label{Eq:capF-Fourier}
F(\zeta)=\frac{1}{2\pi}\int_{-\infty}^\infty dt\, e^{i\zeta t}\, 
\psi_1(t),
\end{equation}
then it follows from Eq.~(\ref{Eq:MSW-2-nu-other}) that $F(\zeta)$
satisfies the first-order differential equation
\begin{equation}
-\zeta^2F(\zeta)-\frac{d}{d\zeta}[-i\zeta F(\zeta)]
+(a_2^2-i)F(\zeta)=0, \nonumber
\end{equation}
where we have omitted all terms from $t=\pm\infty$.
This differential equation simplifies immediately to
\begin{equation}
i\zeta\,\frac{dF(\zeta)}{d\zeta}-(\zeta^2-a_2^2)F(\zeta)=0, \nonumber
\end{equation}
or
\begin{equation}
\label{Eq:capF-d.e.}
\frac{1}{F(\zeta)}\,\frac{dF(\zeta)}{d\zeta}
=\frac{i}{\zeta}(a_2^2-\zeta^2).
\end{equation}
Integration over $\zeta$ gives
\begin{equation}
\label{Eq:capF}
F(\zeta)=\text{const.}\,e^{-i\zeta^2/2}\, \zeta^{ia_2^2}.
\end{equation}

 From the inequality (\ref{Eq:a-positive}), it is seen that 
the function on the right-hand side of Eq.~(\ref{Eq:capF}) has
a singularity at
\begin{equation}
\zeta=0=b_2.
\end{equation}
Therefore the constant in (\ref{Eq:capF}) can take on different
values for $\zeta$ positive and for $\zeta$ negative.
In other words, the differential equation (\ref{Eq:capF-d.e.})
is really two differential equations, one for $\zeta>0$ and
the other for $\zeta<0$, consistent with the fact that the
right-hand side of Eq.~(\ref{Eq:capF-d.e.}) has a singularity
at $\zeta=0$. With this observation, it is natural to define
\addtocounter{equation}{1}
\begin{equation}
F_1(\zeta)
=\begin{cases}
\label{Eq:capF-first case}
e^{-i\zeta^2/2}\,(-\zeta)^{ia_2^2} & \text{for $\zeta<0$}, \\
0                                  & \text{for $\zeta>0$},
\end{cases} \tag{\theequation a}
\end{equation}
and
\begin{equation}
F_2(\zeta)
=\begin{cases}
0                                  & \text{for $\zeta<0$}, \\
e^{-i\zeta^2/2}\,\zeta^{ia_2^2}    & \text{for $\zeta>0$}.
\end{cases} \tag{\theequation b}
\end{equation}
Inverting the Fourier transform (\ref{Eq:capF-Fourier}), this choice 
leads to
\begin{eqnarray}
\label{Eq:psi1-1-2}
\psi_1^{(1)}(t)&=&\int_{-\infty}^0 d\zeta\, e^{-i\zeta t}\,
e^{-i\zeta^2/2}\,(-\zeta)^{ia_2^2}\, , \nonumber \\
\psi_1^{(2)}(t)&=&\int_0^\infty d\zeta\, e^{-i\zeta t}\,
e^{-i\zeta^2/2}\,\zeta^{ia_2^2}\, .
\end{eqnarray}
With the notation (\ref{Eq:b-extras}), these two formulas 
(\ref{Eq:psi1-1-2}) can be written as
\begin{equation}
\psi_1^{(n)}(t)=\int_{b_n}^{b_{n+1}} d\zeta\, e^{-i\zeta t}\,
e^{-i\zeta^2/2}\,|\zeta|^{ia_2^2},
\end{equation}
for $n=1,2$.

It remains to show that both $\psi_1^{(1)}(t)$ and $\psi_1^{(2)}(t)$
are confluent hypergeometric functions of the correct parameters 
and argument.
For this purpose, it is convenient to define
\begin{eqnarray}
\psi_c(t)&=&\int_0^\infty d\zeta\, \cos(\zeta t)
e^{-i\zeta^2/2}\,\zeta^{ia_2^2}, \nonumber \\
\psi_s(t)&=&\int_0^\infty d\zeta\, \sin(\zeta t)
e^{-i\zeta^2/2}\,\zeta^{ia_2^2},
\end{eqnarray}
so that it follows from Eqs.~(\ref{Eq:psi1-1-2}) that
\begin{eqnarray}
\psi_1^{(1)}(t)&=&\psi_c(t)+i\psi_s(t), \nonumber \\
\psi_1^{(2)}(t)&=&\psi_c(t)-i\psi_s(t).
\end{eqnarray}
It is found that
\begin{equation}
\label{Eq:psic-res}
\psi_c(t)=e^{-i\pi/4}\,e^{\pi a_2^2/4}\, 2^{(-1+ia_2^2)/2}\,
\Gamma(\half+\half ia_2^2)\,
\Phi(\half+\half ia_2^2,\half;\half it^2)
\end{equation}
and
\begin{equation}
\label{Eq:psis-res}
\psi_s(t)=-i\,e^{\pi a_2^2/4}\, 2^{ia_2^2/2}\,
\Gamma(1+\half ia_2^2)\,t\,
\Phi(1+\half ia_2^2,\threehalf;\half it^2).
\end{equation}
There are various ways to verify Eqs.~(\ref{Eq:psic-res}) and
(\ref{Eq:psis-res}), including carrying out power series expansions 
in $t$ for the left-hand and right-hand sides.

Finally, we note from Eq.~(7) on p.~257 of reference \cite{Bateman-1}
that
\begin{equation}
\Psi(a,c;x)=\frac{\Gamma(1-c)}{\Gamma(a-c+1)}\,\Phi(a,c;x)
+\frac{\Gamma(c-1)}{\Gamma(a)}\, x^{1-c}\,\Phi(a-c+1,2-c;x).
\end{equation}
Therefore, the results of Sec.~3 and this section are the same.
\section{General values of $N$}
\setcounter{equation}{0}
The procedure of Sec.~4 for $N=2$ can be generalized in 
a straightforward way to larger values of $N$. Indeed, this is the
major advantage over the previously known ones as reviewed in Sec.~3.
This generalization to arbitrary values of $N$ is to be carried out
in this section. Thus, the differential equations (\ref{Eq:psi-Schr})
need to be solved under the constraints (\ref{Eq:sum-b-zero}),
(\ref{Eq:b-ordered}), and (\ref{Eq:a-positive}).

By Eqs.~(\ref{Eq:psi}) and (\ref{Eq:A}), the Eqs.~(\ref{Eq:psi-Schr})
are more explicitly
\begin{equation}
\label{Eq:psi1-5}
i\,\frac{d\psi_1(t)}{dt}=-t\,\psi_1(t)+\sum_{j=2}^N a_j\psi_j(t)
\end{equation}
and, for $k=2,3,4\ldots N$,
\begin{equation}
\label{Eq:psik-5}
\left(i\,\frac{d}{dt}-b_k\right)\psi_k(t)=a_k\psi_1(t).
\end{equation}
In order to get a differential equation for $\psi_1(t)$,
apply the operator
\begin{equation}
\prod_{k=2}^N\left(i\,\frac{d}{dt}-b_k\right) \nonumber
\end{equation}
to Eq.~(\ref{Eq:psi1-5}). By Eq.~(\ref{Eq:psik-5}), this gives
\begin{equation}
\label{Eq:psiN-5}
\left[\prod_{k=2}^N\left(i\,\frac{d}{dt}-b_k\right)\right]
\left(i\,\frac{d}{dt}+t\right)\psi_1(t)
=\sum_{j=2}^N a_j^2
\prod_{\substack{k=2\\ k\ne j}}^N
\left(i\,\frac{d}{dt}-b_k\right)\psi_1(t).
\end{equation}
Eq.~(\ref{Eq:psiN-5}) is an $N$th-order ordinary differential
equation for $\psi_1(t)$; it reduces to Eq.~(\ref{Eq:MSW-2-nu-other})
when $N=2$.

Following Sec.~4, define the Fourier transform $F(\zeta)$ of $\psi_1(t)$
by Eq.~(\ref{Eq:capF-Fourier}), then the first-order differential
equation for $F(\zeta)$ is
\begin{equation}
\left[\prod_{k=2}^N(\zeta-b_k)\right]
\left(\zeta-i\,\frac{d}{d\zeta}\right)F(\zeta)
=\sum_{j=2}^N a_j^2
\prod_{\substack{k=2\\ k\ne j}}^N(\zeta-b_k)F(\zeta),
\end{equation}
or
\begin{equation}
\left(\zeta-i\,\frac{d}{d\zeta}\right)F(\zeta)
=\sum_{j=2}^N\frac{a_j^2}{\zeta-b_j}\,F(\zeta),
\end{equation}
or
\begin{equation}
\label{Eq:capF-d.e.-5}
\frac{1}{F(\zeta)}\,\frac{d F(\zeta)}{d\zeta}
=i\left(-\zeta+\sum_{j=2}^N\,\frac{a_j^2}{\zeta-b_j}\right).
\end{equation}
This Eq.~(\ref{Eq:capF-d.e.-5}) is the generalization of the previous 
Eq.~(\ref{Eq:capF-d.e.}) for $N=2$.
Integration over $\zeta$ gives the generalization of Eq.~(\ref{Eq:capF}):
\begin{equation}
\label{Eq:capF-5}
F(\zeta)=\text{const.}\,e^{-i\zeta^2/2}\, 
\prod_{j=2}^N(\zeta-b_j)^{ia_j^2}.
\end{equation}

 From (\ref{Eq:a-positive}), the function on the right-hand side
of this Eq.~(\ref{Eq:capF-5}) has singularities at
\begin{equation}
\zeta=b_j
\end{equation}
for $j=2,3,\ldots N$.
Therefore, define for $n=1,2,3\ldots N$
\begin{equation}
F_n(\zeta)
\label{Eq:capF-5-sol}
=\begin{cases}
e^{-i\zeta^2/2}\,\prod_{j=2}^N|\zeta-b_j|^{ia_j^2} & 
\text{for $b_n<\zeta<b_{n+1}$}, \\
0                                  & \text{otherwise}.
\end{cases} 
\end{equation}
Inverting the Fourier transform (\ref{Eq:capF-Fourier}) then gives
the desired $N$ linearly independent solutions of the differential
equation (\ref{Eq:psiN-5}) as
\begin{equation}
\label{Eq:psi1-5-n}
\psi_1^{(n)}(t)
=\int_{b_n}^{b_{n+1}} d\zeta\, e^{-i\zeta t}\, e^{-i\zeta^2/2}
\prod_{j=2}^N|\zeta-b_j|^{ia_j^2}
\end{equation}
for $n=1,2,3,\ldots N$.
In both Eq.~(\ref{Eq:capF-5-sol}) and Eq.~(\ref{Eq:psi1-5-n}),
the notation of (\ref{Eq:b-extras}) has been used.
The general solution of (\ref{Eq:psiN-5}) is of course
\begin{equation}
\psi_1(t)=\sum_{n=1}^N C_n\,\psi_1^{(n)}(t)
\end{equation}
with arbitrary constants $C_n$.

The other components of the $\psi(t)$ of (\ref{Eq:psi}) can be easily
obtained also, and the result is
\begin{equation}
\label{Eq:psi-n-sum}
\psi(t)=\sum_{n=1}^N C_n\,\psi^{(n)}(t),
\end{equation}
where
\begin{equation}
\label{Eq:psi-n}
\psi^{(n)}(t)=\int_{b_n}^{b_{n+1}} d\zeta\, e^{-i\zeta t}\, 
e^{-i\zeta^2/2} \left(\prod_{j=2}^N|\zeta-b_j|^{ia_j^2}\right)
\begin{bmatrix}
1 \\ \frac{a_2}{\zeta-b_2} \\ \frac{a_3}{\zeta-b_3} \\ \vdots \\
\frac{a_{N-1}}{\zeta-b_{N-1}} \\ \frac{a_N}{\zeta-b_N}
\end{bmatrix}.
\end{equation}
\section{Limiting Behaviors for Large Distances}
\setcounter{equation}{0}
The next task is to obtain the limiting behaviors of the various
components of the wave function when the distance $x$ is large,
either positive or negative.
In other words, the problem is to find the limiting behaviors of
the $\psi_j^{(n)}(t)$, as given explicitly by Eq.~(\ref{Eq:psi-n}),
both for $t\to-\infty$ and $t\to\infty$, with all the $a$'s and
$b$'s fixed. It is important to remember that these two limits correspond 
to different physical problems, as discussed after Eq.~(\ref{Eq:W11}).

The consideration here will be limited to the part of the asymptotic 
behavior that does not vanish as $t\to\pm\infty$.
This is the physically interesting part. 
There are two possible types of contributions, 
from points of stationary phase and from end points of integration.

\subsection{Points of Stationary Phase}
From Eq.~(\ref{Eq:psi-n}), the points of stationary phase are determined 
by
\begin{equation}
\label{Eq:stat-point-1}
\frac{\partial}{\partial\zeta}(-\zeta t -\half \zeta^2) = 0
\end{equation}
or
\begin{equation}
\label{Eq:stat-point-2}
\zeta  = -t.
\end{equation}
In Eq.~(\ref{Eq:stat-point-1}), the additional phase due to the factor
\begin{equation}
\prod_{j=2}^N |\zeta-b_j|^{i a_j^2} \nonumber
\end{equation}
is not included because the $a_j$ and $b_j$ are all fixed while
$t\to\pm\infty$. Eq.~(\ref{Eq:stat-point-2}) implies that this point
of stationary point is relevant only to:
\begin{itemize}
\item[$\bullet$]
$\psi_1^{(1)}(t)$ as $t\to\infty$, and
\item[$\bullet$]
$\psi_1^{(N)}(t)$ as $t\to-\infty$
\end{itemize}
in view of Eq.~(\ref{Eq:psi-n}).
For example, when $j>1$, $\psi_j^{(1)}(t)$ as $t\to\infty$ and
$\psi_j^{(N)}(t)$ as $t\to-\infty$ both behave as $1/t$ in absolute value
so far as the contribution from this point of stationary phase
(\ref{Eq:stat-point-2}) is concerned.

\subsection{End Points of Integration}
It is seen from Eq.~(\ref{Eq:psi-n}) that, when $k\ge2$, there is an extra
factor of
\begin{equation}
\label{Eq:extra-factor}
\frac{a_k}{\zeta-b_k}
\end{equation}
associated with $\psi_k^{(n)}(t)$.
But the range of integration for this $\psi_k^{(n)}(t)$ as given by
Eq.~(\ref{Eq:psi-n}) is from $b_n$ to $b_{n+1}$.
Therefore, the contribution from these end points of integration 
can lead to a non-zero answer only when the index $k$ appearing
in the expression (\ref{Eq:extra-factor}) agrees with either $n$
or $n+1$. In other words, there are non-zero contributions
as $t\to\pm\infty$ only to $\psi_k^{(k-1)}(t)$ [i.e., $n=k-1$]
and $\psi_k^{(k)}(t)$ [i.e., $n=k$].
These particular components are given by
\begin{eqnarray}
\label{Eq:end-terms}
&&\psi_k^{(k-1)}(t)
=\int_{b_{k-1}}^{b_k} d\zeta\, e^{-i\zeta t}\, e^{-i\zeta^2/2}
\biggl[\prod_{j=2}^{k-1}(\zeta-b_j)^{ia_j^2}\biggr]
\biggl[\prod_{j=k}^{N}(b_j-\zeta)^{ia_j^2}\biggr]\,
\frac{-a_k}{b_k-\zeta}\, , \nonumber \\
&&\psi_k^{(k)}(t)
=\int_{b_{k}}^{b_{k+1}} d\zeta\, e^{-i\zeta t}\, e^{-i\zeta^2/2}
\biggl[\prod_{j=2}^{k}(\zeta-b_j)^{ia_j^2}\biggr]
\biggl[\prod_{j={k+1}}^{N}(b_j-\zeta)^{ia_j^2}\biggr]\,
\frac{a_k}{\zeta-b_k} . \nonumber \\
\end{eqnarray}
These Eq.~(\ref{Eq:end-terms}) are exact.

Since the important contributions come from the vicinity of $\zeta=b_k$,
all the $\zeta$'s in Eq.~(\ref{Eq:end-terms}) can be replaced 
approximately by $b_k$ except in the factors $e^{-i\zeta t}$,
$b_k-\zeta$, and $\zeta-b_k$.
Therefore
\begin{eqnarray}
\psi_k^{(k-1)}(t)
&\sim& e^{-ib_k^2/2}
\biggl[\prod_{\substack{j=2\\ j\ne k}}^{N}
|b_j-b_k|^{ia_j^2}\biggr]
(-a_k) \int^{b_k} d\zeta\, e^{-i\zeta t}\, (b_k-\zeta)^{-1+ia_k^2}\, , 
\nonumber \\
\psi_k^{(k)}(t)
&\sim& e^{-ib_k^2/2}
\biggl[\prod_{\substack{j=2\\ j\ne k}}^{N}
|b_j-b_k|^{ia_j^2}\biggr]
a_k \int_{b_k} d\zeta\, e^{-i\zeta t}\, (\zeta-b_k)^{-1+ia_k^2}\, ,
\end{eqnarray}
or
\begin{eqnarray}
\psi_k^{(k-1)}(t)
&\sim& e^{-ib_k^2/2}\, e^{-ib_k t}
\biggl[\prod_{\substack{j=2\\ j\ne k}}^{N}
|b_j-b_k|^{ia_j^2}\biggr]
(-a_k) \int_0^\infty dx \, e^{ixt}\,x^{-1+ia_k^2}\, ,
\nonumber \\
\psi_k^{(k)}(t)
&\sim& e^{-ib_k^2/2}\, e^{-ib_k t}
\biggl[\prod_{\substack{j=2\\ j\ne k}}^{N}
|b_j-b_k|^{ia_j^2}\biggr]
a_k \int_0^\infty dx\, e^{-ixt}\, x^{-1+ia_k^2}.
\end{eqnarray}
This integral can be evaluated exactly in terms of the gamma function.

\subsection{Results}
\begin{figure}[htb]
\refstepcounter{figure}
\label{Fig:figure1}
\addtocounter{figure}{-1}
\begin{center}
\setlength{\unitlength}{1cm}
\begin{picture}(8.0,6.0)
\put(0.5,0.0){
\mbox{\epsfysize=8cm\epsffile{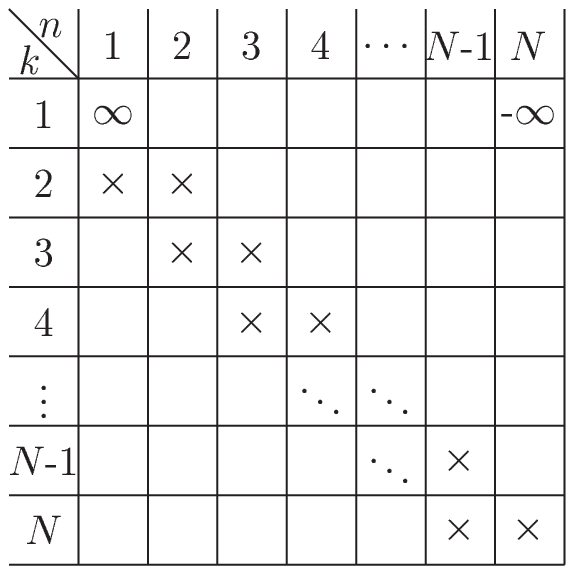}}}
\end{picture}
\caption{
Table of non-vanishing components of $\psi_k^{(n)}(t)$ as $t\to\pm\infty$.
A cross means that the component is non-vanishing both for $t\to-\infty$
and $t\to+\infty$; the symbol $\infty$ means for $t\to+\infty$ only,
and $-\infty$ means for $t\to-\infty$ only.}
\end{center}
\end{figure}

Fig.~1 shows which ones of the various $\psi_k^{(n)}(t)$ have
non-vanishing behaviors for $t\to-\infty$ and $t\to\infty$.

These non-vanishing behaviors are:

\noindent
For $t$ positive and large,
\begin{eqnarray}
\label{Eq:res-pos-1-1}
\psi_1^{(1)}(t)
&\sim& \sqrt{2\pi}\, e^{-i\pi/4}\, t^{i\alpha}\, e^{it^2/2}, \\
\label{Eq:res-pos-km1}
\psi_k^{(k-1)}(t)
&\sim& e^{-ib_k^2/2}\, e^{-ib_k t}
\biggl[\prod_{\substack{j=2\\ j\ne k}}^{N}
|b_j-b_k|^{ia_j^2}\biggr]
(-a_k) e^{-\pi a_k^2/2}\, \Gamma(ia_k^2) t^{-ia_k^2}, \\
\label{Eq:res-pos-k}
\psi_k^{(k)}(t)
&\sim& e^{-ib_k^2/2}\, e^{-ib_k t}
\biggl[\prod_{\substack{j=2\\ j\ne k}}^{N}
|b_j-b_k|^{ia_j^2}\biggr]
a_k\, e^{\pi a_k^2/2}\, \Gamma(ia_k^2) t^{-ia_k^2};
\end{eqnarray}
while, for $t$ negative and large,
\begin{eqnarray}
\label{Eq:res-neg-1-N}
\psi_1^{(N)}(t)
&\sim& \sqrt{2\pi}\, e^{-i\pi/4}\, |t|^{i\alpha}\, e^{it^2/2}, \\
\label{Eq:res-neg-km1}
\psi_k^{(k-1)}(t)
&\sim& e^{-ib_k^2/2}\, e^{-ib_k t}
\biggl[\prod_{\substack{j=2\\ j\ne k}}^{N}
|b_j-b_k|^{ia_j^2}\biggr]
(-a_k) e^{\pi a_k^2/2}\, \Gamma(ia_k^2) |t|^{-ia_k^2}, \\
\label{Eq:res-neg-k}
\psi_k^{(k)}(t)
&\sim& e^{-ib_k^2/2}\, e^{-ib_k t}
\biggl[\prod_{\substack{j=2\\ j\ne k}}^{N}
|b_j-b_k|^{ia_j^2}\biggr]
a_k\, e^{-\pi a_k^2/2}\, \Gamma(ia_k^2) |t|^{-ia_k^2}.
\end{eqnarray}
All the other components approach zero as $t\to\infty$ and as
$t\to-\infty$. 
In the asymptotic formulas (\ref{Eq:res-pos-1-1}) and 
(\ref{Eq:res-neg-1-N}), $\alpha$ is the quantity
\begin{equation}
\alpha=\sum_{j=2}^N a_j^2 .
\end{equation}
In the formulas (\ref{Eq:res-pos-km1}), (\ref{Eq:res-pos-k}), 
(\ref{Eq:res-neg-km1}) and (\ref{Eq:res-neg-k}),
\begin{equation}
2\le k \le N,
\end{equation}
where $N$ as always is the number of neutrino species.
\section{Discussion}
\setcounter{equation}{0}
When we started to investigate the MSW differential equations
for three neutrino species in the case of the linear electron
density, we were mostly interested in various possibilities of finding
approximate solutions.
Therefore, it was quite a surprise to us that these coupled
differential equations can be solved exactly not only for three,
but also for any number of neutrino species.

In the work of Wolfenstein, Mikheyev, Smirnov \cite{MSW} and others
\cite{two-nu} on the sun taking into account two species of neutrinos,
it has been found that most of the effect takes place in a fairly
narrow region around a particular value of the electron density.
Because of this, it is quite accurate to use a linear approximation
to the electron density.

For more than two species of neutrinos, it is no longer true in general
that there is a narrow region for most of the activity.
Nevertheless, there are a number of circumstances where this is true.
However, the conditions for this to hold has not yet been studied
systematically. This is one direction for future work.

Under the assumption of the electron density being a linear function
of distance, the exact, general solution of the MSW differential
equation is given by Eqs.~(\ref{Eq:psi-n-sum}) and (\ref{Eq:psi-n}).
This solution is in the form of a number of single integrals.
When the number of neutrino species is more than 2, these integrals
cannot be evaluated in terms of known functions, and therefore
their properties need to be investigated.
A small step in this direction has been taken in Sec.~6,
where the asymptotic behaviors of these integrals have been evaluated
for large distances but with all the $a$'s and $b$'s held fixed.
It is believed that, in so far as this case of linear electron density
is applicable to the physically interesting case of solar neutrinos,
the asymptotic evaluation of Sec.~6 is far from being sufficient.
It is more likely that not only the distance, but also some of
the parameters, the $a$'s and $b$'s, are large.
This is a second direction for future work.

\bigskip

\leftline{\bf Acknowledgments}
\par\noindent
We are greatly indebted to Dr.\ Conrad Newton
for collaboration at the early stage of this work.
One of us (TTW) thanks the Theory Division at CERN for its kind hospitality.

\end{document}